\documentstyle[12pt,epsf]{article}    
\newcommand\beq{\begin{eqnarray}}
\newcommand\eeq{\end{eqnarray}}

\newcommand\la{\langle}
\newcommand\ra{\rangle}

\begin{document}

\setlength{\baselineskip}{0.3in}



\vskip 1.0cm

\Large
\centerline{\bf Bag model prediction for the nucleon's chiral-odd}
\centerline{\bf twist-3 distribution $h_L(x,Q^2)$ at high $Q^2$}
\normalsize

\vspace{1.5cm}
\centerline{Y. Kanazawa and Yuji Koike}

\centerline{\it Graduate School of Science and Technology, 
Niigata University,
Ikarashi, Niigata 950-21, Japan}

\vspace{3cm}

\centerline{\bf Abstract}
We study the $Q^2$ evolution of the nucleon's chiral-odd twist-3
distribution $h_L(x,Q^2)$ starting from the MIT bag model calculation.
A simple GLAP equation for $h_L(x,Q^2)$ obtained
at large $N_c$ is used for the $Q^2$ evolution.  The correction
due to the finite value of $N_c$ is $O(1/N_c^2)\sim 10$ \% level.
It turns out that the twist-3 contribution to $h_L(x,Q^2)$
is significantly reduced at $Q^2=10$ GeV$^2$ 
in contrast to the $g_2(x,Q^2)$ case.  This is due to the fact that
the corresponding anomalous dimension for $h_L$ is larger than that for
$g_2$ at small $n$ (spin).




\newpage

The EMC measurement of the nucleon's $g_1$ structure function\,\cite{EMC}
inspired lots of theoretical activities on the 
nucleon's spin-structure functions in general as well as
more precision measurements of $g_1$\,\cite{EMC2}.
These structure functions provide us with a rich source of
information about the spin distributions of quarks and gluons
inside the nucleon.  
Jaffe and Ji\,\cite{JJ} discussed 
general features of the quark distributions
of the nucleon and relevant places where they can be measured. 
The nucleon has three independent twist-2 quark distributions, $f_1(x,Q^2)$ 
 (spin-average),
$g_1(x,Q^2)$ (helicity asymmetry), $h_1(x,Q^2)$
(helicity flip), and three independent twist-3 quark
distributions $e(x,Q^2)$, $g_2(x,Q^2)$, $h_L(x,Q^2)$.
Twist-2 distributions have a simple
parton model interpretation and contribute to various hard processes
in the leading order with respect to $1/\sqrt{Q^2}$. ($Q$ is the hard momentum
of the external hard probe.)  On the other hand, the twist-3 distributions
represent complicated quark-gluon correlations in the nucleon, and 
is generally difficult to be measured, since they are often hidden 
behind the leading twist-2 contributions.  However, $g_2$ and $h_L$
can be measured in the absence of the leading twist-2 contributions
through the proper asymmetries in the polarized deep inelastic
scattering and the polarized Drell-Yan process, respectively\,\cite{Jg2,JJ}.
In this sence, they are interesting higher twist distribution functions.
In fact, E143 collaboration\,\cite{E143} presented a first nonzero data for
$g_2$, which anticipates a forthcoming significant 
progress in twist-3 physics.

So far accumulated experimental data on $f_1$ and $g_1$ 
allowed us to parametrize
in the next-to-leading order for $f_1$\,\cite{GRV} 
and for $g_1$\,\cite{GRV2}.  
But nothing is known
about the actual shape of $h_1$, $g_2$ and $h_L$
except some guess by the
bag model calculations\,\cite{JJ,JJ91,St}. (Since there is no 
practical way 
of isolating $e$, it will not be considered in this work.)
The bag model has been reasonably successful in describing various properties
of hadrons\,\cite{bag}, and has been applied to calculate the structure 
functions\,\cite{JJ91,St,J75,BM,STL,SST}.
Since the bag model is a low energy effective hadron model, 
its prediction for the structure functions
has to be
evolved to higher scale to confront experimental data.
After the $Q^2$ evolution, it could approximately reproduce the 
valence parts of $f_1$ and $g_1$.
The purpose of this short note is to present the first and a rough
estimate of the magnitude of $h_L(x,Q^2)$ at high $Q^2$ staring from the
bag model calculation.  
We are especially interested in the speed of the $Q^2$-evolution
of $h_L$ compared with that of 
$g_2$ and the chiral-odd twist-2 distribution $h_1$.
Since it is not our purpose here to
construct a more realistic model, we shall not persue
the projection method 
to restore the translational invariance as was tried
in \cite{St,BM,STL,SST}.
We refer those attempts to future studies.

We first recall the definition of the quark distributions 
in our interest\,\cite{JJ}:
\beq
& &\int{d\lambda \over 2\pi} e^{i\lambda x}\la PS|\bar{\psi}
(0)\gamma_\mu \gamma_5 \psi(\lambda n)|_Q |PS \ra  
\nonumber\\ & & \qquad=
2\left[ g_1(x,Q^2)p_\mu (S\cdot n) + g_T(x,Q^2) 
S_{\perp\mu} + M^2 g_3(x,Q^2)S\cdot n
n_\mu \right],
\label{eqg12}\\
& &\int{d\lambda \over 2\pi} e^{i\lambda x}\la PS|\bar{\psi}
(0)\sigma_{\mu\nu} i\gamma_5 \psi(\lambda n)|_Q |PS \ra 
=2\left[ h_1(x,Q^2)\left( S_{\perp\mu}p_\nu - S_{\perp\nu}p_\mu \right)/M
\right.\nonumber\\
& &\qquad\qquad \left.
+h_L(x,Q^2)M \left( p_\mu n_\nu - p_\nu n_\mu \right) (S\cdot n)
+ h_3(x,Q^2)M 
\left(S_{\perp\mu}n_\nu - S_{\perp\nu}n_\mu \right) \right],
\label{eqh12}
\eeq
where $|PS\ra$ denotes the nucleon (mass $M$) state with the four momentum
$P$ and the spin $S$, and the two light-like vectors, $p$ and $n$,
are introduced
by the relation $P^\mu =p^\mu + {M^2\over 2}n^\mu$, $p\cdot n=1$,
$p^2=n^2=0$.  
For the nucleon moving in the $z$-direction, 
$p={{\cal P}\over \sqrt{2}}(1,0,0,1)$ and 
$n={ 1 \over \sqrt{2}{\cal P}}(1,0,0,-1)$.  ${\cal P}\to \infty$
corresponds to the infinite momentum frame and ${\cal P}=M/\sqrt{2}$
corresponds to the nucleon's rest frame.   
$S^\mu$ is decomposed as $S^\mu=(S\cdot n) p^\mu + (S\cdot p) n^\mu + 
S_\perp^\mu$.  In (\ref{eqg12}) and (\ref{eqh12}), 
lightcone gauge, $n\cdot A \sim A^+ =0$, was employed.
The above distribution functions $g_{1,T}$ ($g_T=g_1+g_2$)
and $h_{1,L}$ {\it etc} 
are defined for each quark flavor
$\psi = \psi^a$ ($a=u, d, s,...$) and have support $-1< x <1$\,\cite{J83}.
The replacement
$\psi^a \to C\bar{\psi^a}^T$, $\bar{\psi^a}\to -\psi^a C^{-1}$
defines the anti-quark distributions $\bar{g^a_{1,T}}(x)$ {\it etc}
for each quark distribution $g^a_{1,T}(x)$ {\it etc}.
They are related as $g_{1,T}^a(-x)=\bar{g_{1,T}^a}(x)$,
$h_{1,L}^a(-x)=-\bar{h_{1,L}^a}(x)$.
For the polarized deep inelastic scattering, physically measurable
structure functions are the combination
$\sum_a e_a^2 (g_{1,T}^a(x)+\bar{g_{1,T}^a}(x))$ with 
the Bjorken $x$ ($0<x<1$) and the electric
charge of a (anti-)quark flavor $a$, $e_a$ . 
Here and below, we often suppress the explicit $Q^2$ dependence
of the distributions.

The $Q^2$ dependence of these structure functions 
is calculable in perturbative QCD.  
The twist-2 distributions, $g_1$ and $h_1$,
obey simple GLAP equations\,\cite{GLAP}.  
On the other hand, the $Q^2$ dependence
of the twist-3 distributions, $g_2$ and $h_L$,
is quite sophisticated because the number of
quark-gluon-quark operators increases with the moments (or spin).
The calculation of the one-loop anomalous dimension matrix 
for all the twist-3 distributions has been completed\,\cite{SV,BKL,KT,KN,BM2},
and
an analogue of GLAP 
equation relevant to describe $Q^2$ evolution 
of the whole $x$ dependent
distributions 
has also been derived in \cite{BKL,BM2}.
These equations for the twist-3 distributions
are the evolution equation for the corresponding parent distributions
and is not convenient for practical applications.
However, there is a very useful news
for physicists working on higher twist effects.
It has been proved that at large $N_c$, the 
$Q^2$ evolution of all the twist-3 distributions 
is described by simple GLAP
equations with slightly different forms for the
anomalous dimensions from the twist-2 distributions\,\cite{ABH,BBKT,KN}:
\footnote{
In a recent work\cite{AVB}, it has also been shown that the
same simplification at large $N_c$ occurs for
the $Q^2$ dependence of all the twist-3 fragmentation functions.}
The $Q^2$ evolution (for $g_2$, only for nonsinglet piece) is given by
\beq
{\cal M}_n\left[ \tilde{g}_2(Q^2)\right]
=L^{\gamma_n^g/b_0}{\cal M}_n\left[ \tilde{g}_2(\mu^2)\right],
\label{eq1}\\
{\cal M}_n\left[ \tilde{h}_L(Q^2)\right]
=L^{\gamma_n^h/b_0}{\cal M}_n\left[ \tilde{h}_L(\mu^2)\right],
\label{eq2}\\
{\cal M}_n\left[ e(Q^2)\right]
=L^{\gamma_n^e/b_0}{\cal M}_n\left[ e(\mu^2)\right],
\label{eq3}
\eeq
where
${\cal M}_n[g(Q^2)] \equiv \int_{-1}^1dx\,x^ng(x,Q^2)$, $L\equiv
\alpha_s(Q^2)/\alpha_s(\mu^2)$, $b_0={11\over 3}N_c - {2\over 3}N_f$.
$\widetilde{g}_2$ and $\widetilde{h}_L$ denote the twist-3
parts of $g_2$ and $h_L$, respectively.
The corresponding anomalous dimensions in (\ref{eq1})-
(\ref{eq3}) are given by
\beq
\gamma_n^g= 2N_c \left( S_n -{1 \over 4} + {1 \over 2(n+1) } \right),
\label{eq4}\\
\gamma_n^h= 2N_c \left( S_n -{1 \over 4} + {3 \over 2(n+1) } \right),
\label{eq5}\\
\gamma_n^e= 2N_c \left( S_n -{1 \over 4} - {1 \over 2(n+1) } \right),
\label{eq6}
\eeq
with
$S_n= \sum_{j=1}^n{1\over j}$.
Furthermore, these anomalous dimensions are the lowest eigenvalues
of the anomalous dimension matrices at large $N_c$.
Since these relations are obtained by a mere replacement $C_F=(N_c^2-1)/2N_c
\rightarrow N_c/2$ in the complete one-loop anomalous dimension matrices
at finite $N_c$, the correction due to the finite 
value of $N_c$ is $O(1/N_c^2) \sim 10\%$
level, which is sufficient for practical application.
The essential ingredient in (\ref{eq1})-(\ref{eq6})
is that a knowlegde on $g_2(x)$, $h_L(x)$ and $e(x)$
at one scale is sufficient to predict them at an arbitrary scale,
which is not the case at finite $N_c$.
This fact provides us with a useful framework to confront
experimental data at various $Q^2$ of the twist-3 distribution. 
In fact (\ref{eq1}) and (\ref{eq4}) were used to predict the shape of
$g_2$ at high $Q^2$ starting from the bag model calculation\,\cite{St}.
A more favorable feature of $h_L$ compared with $g_2$ is
that $h_L$ does not mix with gluon distributions 
owing to its chiral-odd
nature.  Therefore $Q^2$ evolution for $h_L$ and $e$ is given
by (\ref{eq2}), ({\ref{eq3}), ({\ref{eq5}) and (\ref{eq6})
even for the flavor singlet piece and thus we can get more 
reliable and accurate form in the small $x$ region compared to $g_2$.
This work is devoted to the study of $Q^2$ evolution of $h_L$
with (\ref{eq2}) and (\ref{eq5}).

In the rest frame of the nucleon, one can conveniently calculate
the above distributions using the MIT bag model.  The result for 
$h_1$ and $h_L$ with one quark flavor in the
nucleon is given as\,\cite{JJ}
\beq
h_1(x)&=&{ \omega MR \over 2\pi (\omega -1) j_0^2(\omega)}
\int_{|y_{min}|}^\infty ydy \left[ t_0(\omega,y)^2 +
2 t_0(\omega,y)t_1(\omega,y){y_{min} \over y} + 
t_1(\omega,y)^2 \left( { y_{min} \over y } \right)^2 \right],
\nonumber\\
\label{bagh1}\\
h_L(x)&=&{ \omega MR \over 2\pi (\omega -1) j_0^2(\omega)}
\int_{|y_{min}|}^\infty ydy \left[ t_0(\omega,y)^2 
- t_1(\omega,y)^2 \left( 2\left( { y_{min} \over y } \right)^2 
-1 \right) \right].
\label{baghl}
\eeq
Here $t_l$ is given by
\beq
t_l(\omega,y)=\int_0^1dz\,z^2 j_l(\omega z) j_l (yz),
\eeq
where $j_l$ is the $l$-th order spherical Bessel function, and
$\omega$ is determined by the relation
${\rm tan}\,\omega = -\omega/(\omega-1)$.  For the lowest energy mode, 
$\omega=2.04$.   $y_{min}$ is defined as $y_{min}=MRx-\omega$
with the bag radius $R$ determined by the relation $MR=4\omega$.
$h_L$ is decomposed into the twist-2 piece which can be expressed 
in terms of $h_1$ and a purely twist-3 piece $\widetilde{h}_L$ as
\beq
h_L(x)= \left\{
\begin{array}{ll}
\displaystyle{
2x\int_x^1dy {h_1(y) \over y^2} + \widetilde{h}_L(x)}, & \qquad 
0<x<1\\[0.5cm]
\displaystyle{
-2x\int_{-1}^xdy {h_1(y) \over y^2} + \widetilde{h}_L(x)}. & \qquad 
-1<x<0
\end{array}
\right.
\label{hlww}
\eeq
The bag model prediction
above has to be regarded as a distribution at some
low energy scale $Q^2 = \mu_{bag}^2 \leq 1$ GeV$^2$.
For $h_1$, we regard (\ref{bagh1}) as a valence distribution
at this low energy scale.

In order to evolve the bag model prediction for
$h_L$ from $\mu_{bag}^2$ to $Q^2$
according to (\ref{eq2}), we used a method in \cite{GLY}.
For the moment, we symbolically represent $h_{1,L}(x)$
by $h(x)$.  If one defines $h_\pm (x)=h(x)\pm h(-x)
=h(x)\mp \bar{h}(x)$,
the even (odd) moments of $h_{+}(x)$ ($h_{-}(x)$) on the interval $[0,1]$
agree with ${\cal M}_n[h]$, whose $Q^2$ evolution is given by
(\ref{eq2}) and (\ref{eq5}) and its analogue.
We assume $Q^2$ evolution of all the moments of 
$h_{+}(x)$ and $h_{-}(x)$ on $[0,1]$ is described by the same anomalous
dimensions as was often assumed
to describe $Q^2$ evolution of $f_1(x,Q^2)$ and $g_1(x,Q^2)$, 
and construct
$h(x,Q^2)$ on $[-1,1]$.  This is equivalent to assume that 
the $Q^2$ dependence of the moments of
$h(x,Q^2)$ on $[0,1]$ and $[-1,0]$ are separately governed
by the same anomalous dimension in (\ref{eq5}), which is a sufficient
condition to satisfy (\ref{eq2}).
To describe the method in \cite{GLY}, we
introduce Bernstein
polynomial defined by
\beq
b^{(N,k)}(x)=(N+1) { n \choose k } x^k(1-x)^{N-k} = { (N+1)! \over k! }
\sum_{l=0}^{N-k} { (-1)^l x^{k+l} \over l! (N-k-l)! },
\eeq
and note that it satisfies the relation
\beq
\lim_{{\scriptstyle N,k\to \infty} \atop 
{\scriptstyle k/N\to x} } b^{(N,k)}(y) = \delta (x-y)
\label{bern}
\eeq
for $0<x, y <1$.
Using (\ref{bern}), (\ref{eq2}) and (\ref{eq5}), we get
\beq
\widetilde{h}_L(x,Q^2) &=& \lim_{{\scriptstyle N,k\to \infty} \atop 
{\scriptstyle k/N\to x} } { (N+1)! \over k! } \sum_{l=0}^{N-k}
{ (-1)^l \over l!(N-k-l)! } \int^1_{0}dy\,y^{k+l}\widetilde{h}_L(y,Q^2)
\nonumber\\
&=& 
\lim_{{\scriptstyle N,k\to \infty} \atop 
{\scriptstyle k/N\to x} } { (N+1)! \over k! } \sum_{l=0}^{N-k}
{ (-1)^l \over l!(N-k-l)! }L^{\gamma_{k+l}^h/b_0} 
\int^1_{0}dy\,y^{k+l}\widetilde{h}_L(y,\mu^2).
\label{bernmom}
\eeq
Since the summation over $l$ in (\ref{bernmom}) oscillates due to the factor
$(-1)^l$, the direct use of (\ref{bernmom}) is not 
convenient.  To avoid this difficulty
we shall utilize the following procedure.
Expand $L^{\gamma_n^h/b_0}$ as
\beq
L^{\gamma_n^h/b_0}= a(L)\sum_{i=0}{ C_i(L) \over (n+p)^{i+\rho(L)} },
\label{expansion}
\eeq
where $a(L)$, $C_i(L)$ and $p$ are the constants determined below. 
Then (\ref{bernmom}) is rewritten as
\beq
h(x,Q^2)&=&\int_x^1 {dy \over y} b(x,y; Q, \mu)h(y,\mu),
\nonumber\\
b(x,y;Q,\mu) &\equiv& a(L) \left( {x\over y}\right)^{p-1}
\sum_{i=0}\left( {\rm ln}{y\over x} \right)^{i+\rho-1} 
{C_i \over \Gamma(i+\rho) },
\label{kernel}
\eeq
where we have used the relation
\beq
\lim_{{\scriptstyle N,k\to \infty} \atop 
{\scriptstyle k/N\to x} }{ (N+1)! \over k! } \sum_{l=0}^{N-k}
{ (-1)^l \over l!(N-k-l)! } { y^{k+l} \over (k+l+p)^{i+\rho }}
={ \theta(y-x) \over \Gamma(i+\rho) y}\left({x\over y}\right)^{p-1}
\left( {\rm ln}{y\over x} \right)^{i+\rho-1}.
\eeq
The expansion (\ref{expansion}) can be obtained
by applying the following asymptotic expansion
to $\gamma_n^h$ in (\ref{eq5}): 
\beq
S_{n+1} \sim \gamma_E + \ln (n+1) + { 1 \over 2(n+1)} -
\sum_{k=1}^\infty {B_{2k} \over 2k(n+1)^{2k}}
\label{asymp}
\eeq
where $\gamma_E = 0.577...$ is the Euler constant and $B_{2k}$'s are
the Bernoulli numbers. 
This procedure gives $p=1$ and the coefficients $C_i$ in (\ref{expansion}).
(See \cite{GLY,STL} for the details.)  
We have used first four terms (i.e. up
to $B_{8}$) in the expansion (\ref{asymp}), which
gives enough precision.

Next we need to determine the bag scale $\mu_{bag}$.
Phenomenological values for $\mu_{bag}^2$ 
adopted in the previous studies scatters below 1 GeV$^2$:  
Jaffe and Ross\,\cite{JR80} extracted
$\mu_{bag}^2=0.75$ GeV$^2$ from the 5-th moment of a $F_3$ data.
Schreiber {\it et al.}\,\cite{STL} took $\mu_{bag}^2=0.25$ GeV$^2$ to
reproduce approximate shape of $F_2$ at $Q^2=10$ GeV$^2$.
More recently, Stratmann determined $\mu_{bag}^2=0.081$ GeV$^2$
by comparing the second moment
(momentum sum rule) of the bag model prediction with that of
the valence distribution determined by Gl\"{u}ck {\it et al.}\,\cite{GRV}.
\footnote{Comparison with other parton distributions in \cite{GRV}
gives almost the same numbers for $\mu_{bag}$.}
Our purpose here is to see how $h_L$ evolves compared with
other distributions such as $h_1$ and $g_2$, and therefore
we shall show the results for two values of $\mu_{bag}^2$,
$\mu_{bag}^2=0.081, 0.25$ GeV$^2$, 
for future references.  For other parameters, we set $N_f=3$, 
$\Lambda_{\rm QCD}=0.232$ GeV in $\alpha_s(Q^2)$.

Figure 1 (a) and (b) show the results for $xh_1(x,Q^2)$
and $xg_1(x,Q^2)$ at
$Q^2=10$ GeV$^2$ with two values of
$\mu_{bag}^2$ together with the bag calculations.
For the $Q^2$ evolution, we have used the anomalous
dimension for $h_1$ calculated in \cite{AM,KT}.
$g_1(x,Q^2)$ at $Q^2=10$ GeV$^2$
is strongly peaked in the small $x$ region, since
the anomalous dimension for ${\cal M}_0[g_1]$
is zero while it is 4/3 for ${\cal M}_0[h_1]$.
(If one plots $h_1$ and $g_1$ instead of $xh_1$ and $xg_1$,
this feature is more conspicuous.)
Figure 2 shows the bag calculation for $h_L$
decomposed into the twist-2 and -3 contributions\,\cite{JJ}\,((a)) 
and $h_L$ evolved to
$Q^2=10$ GeV$^2$ with $\mu^2_{bag}=0.081$ GeV$^2$ ((b)) and
$\mu^2_{bag}=0.25$ GeV$^2$ ((c)).   In Fig. 3, we plot only
the twist-3 piece of $h_L$, $\widetilde{h}_L$, taken from Fig. 2
to see how it evolves with $Q^2$.
These graphs show clearly that at higher $Q^2$ 
the contribution from $\widetilde{h}_L(x,Q^2)$ is significantly reduced
and $h_L(x,Q^2)$ is
dominated by the twist-2 contribution.  Although our calculation
starts from the bag model prediction, this tendency can be taken as
model independent.   Comparison of Fig. 3 and Fig. 1 shows
that $\widetilde{h}_L$ evolves faster than $h_1$ as is expected
from the magnitudes of the anomalous dimensions\,\cite{KT}.
For comparison we have also shown $g_2(x,Q^2)$ 
in Fig. 4.
Since this distribution is accessible in the polarized DIS,
we plotted the combination $g_2(x,Q^2)+\bar{g}_2(x,Q^2)
=g_2(x,Q^2)+g_2(-x,Q^2)$ for one quark-flavor with unit charge.
At the bag scale, twist-3 contribution $\widetilde{g}_2$
is comparable to the twist-2 contributin as in the case of
$h_L$.  This feature more or less survives even at $Q^2=10$ GeV$^2$
in contrast to the $h_L$ case.  This is because 
$\gamma_n^h > \gamma_n^g$ especially at small $n$
and hence $Q^2$ evolution of 
$\widetilde{h}_L$ in the small
$x$ region is faster than that of $\widetilde{g}_2$.
As was stated before, flavor-singlet part of $g_2$ mixes with
the gluon distribution and  $Q^2$ evolution of 
singlet $\widetilde{g}_2$ is not given by (\ref{eq1}).  Singlet
$\widetilde{g}_2$ is probably more enhanced at small $x$ region.
If the bag model gives a good description even for
the twist-3 distribution $h_L$, our present study indicates
that it will be extremely difficulte to extract $\widetilde{h}_L(x,Q^2)$
at high $Q^2$.  On the other hand, if future experiments show 
$\widetilde{h}_L(x,Q^2)$ is still sizable at high $Q^2$, it means
that the naive bag model calculation is not suitable to describe
quark-gluon correlation 
represented by $h_L$ in the nucleon.   In any case, it is 
very interesting
to confirm these general features in the future collider experiments.



\newpage
\Large

\centerline{\bf Figure captions}

\normalsize

\vskip 1cm
\noindent
{\bf Fig. 1} (a) Bag model calculation for $xh_1(x)$ (dash-dot line) and 
$h_1(x,Q^2=10\ {\rm GeV}^2)$ with $\mu_{bag}^2$=0.081 GeV$^2$
(solid line) and 0.25 GeV$^2$ (dashed line).
(b) The same as (a) but for for $xg_1(x,Q^2)$.

\vskip 0.5cm

\noindent
{\bf Fig. 2} (a) Bag model calculation for $h_L(x)$ (solid line)
decomposed into
the twist-2 (dashed line) and the twist-3 (dash-dot line) contributions.
(b) $h_L(x, Q^2)$ at $Q^2=10$ GeV$^2$ with $\mu^2_{bag}=0.081$ GeV$^2$
decomposed into the twist-2 and twist-3 contributions.
(c) The same as (b) but with $\mu^2_{bag}=0.25$ GeV$^2$.

\vskip 0.5cm

\noindent
{\bf Fig. 3} Twist-3 contribution to $h_L(x, Q^2)$ at the bag scale 
(dash-dot line)
and $Q^2=10$ GeV$^2$ with $\mu^2_{bag}=$0.081 GeV$^2$ (solid line)
and 0.25 GeV$^2$ (dashed line).

\vskip 0.5cm

\noindent
{\bf Fig. 4} (a) Bag model calculation for $g_2(x)$ decomposed into
the twist-2 ($g_2^{WW}(x)$) 
and the twist-3 ($\widetilde{g}_2(x)$) contributions.  
(b) $g_2(x, Q^2)$ at $Q^2=10$ GeV$^2$ with $\mu^2_{bag}=0.081$ GeV$^2$.
(c) The same as (b) but with $\mu^2_{bag}=0.25$ GeV$^2$.


\begin{figure}[p]
\epsfbox{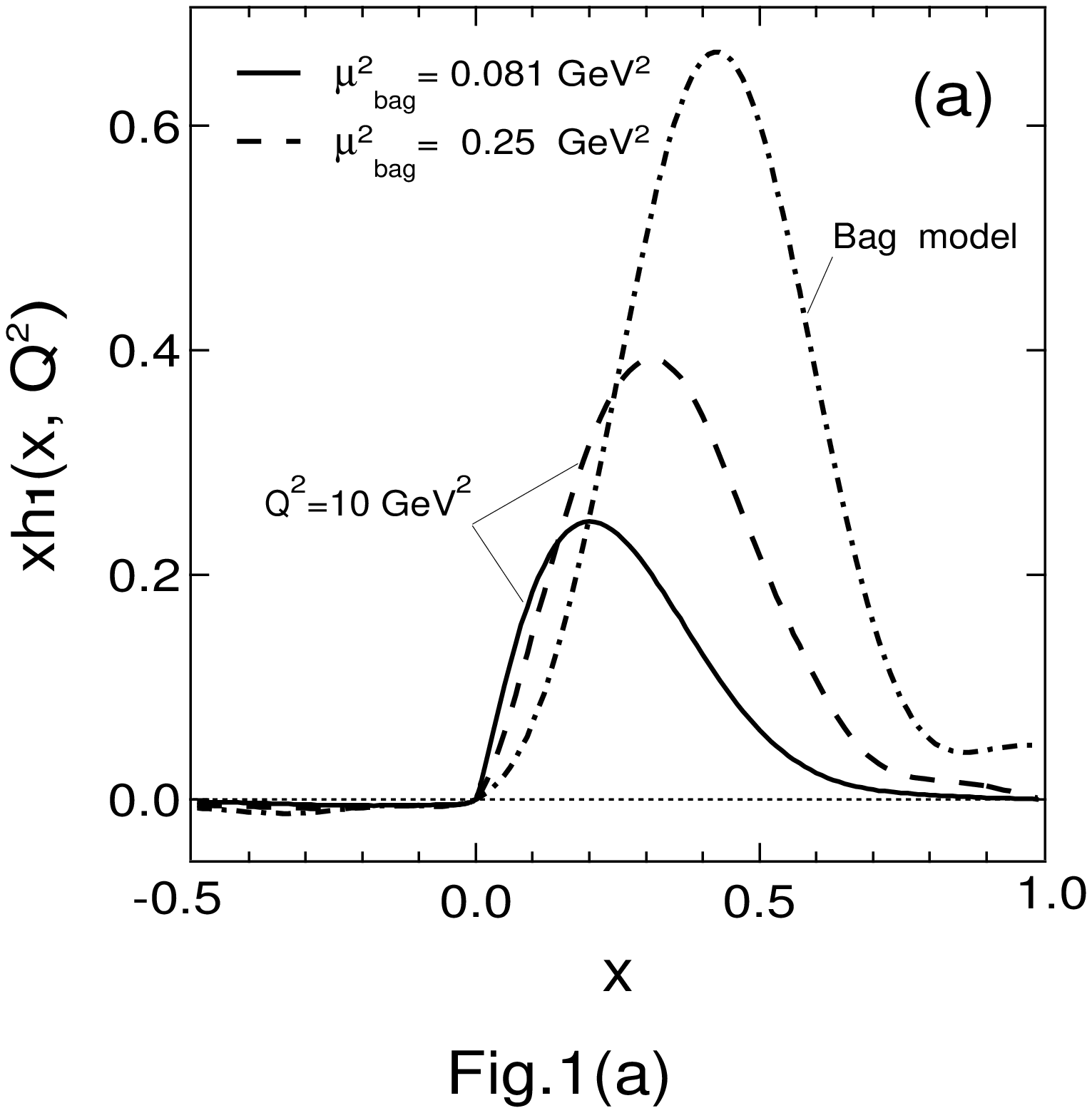}
\end{figure}

\begin{figure}[p]
\epsfbox{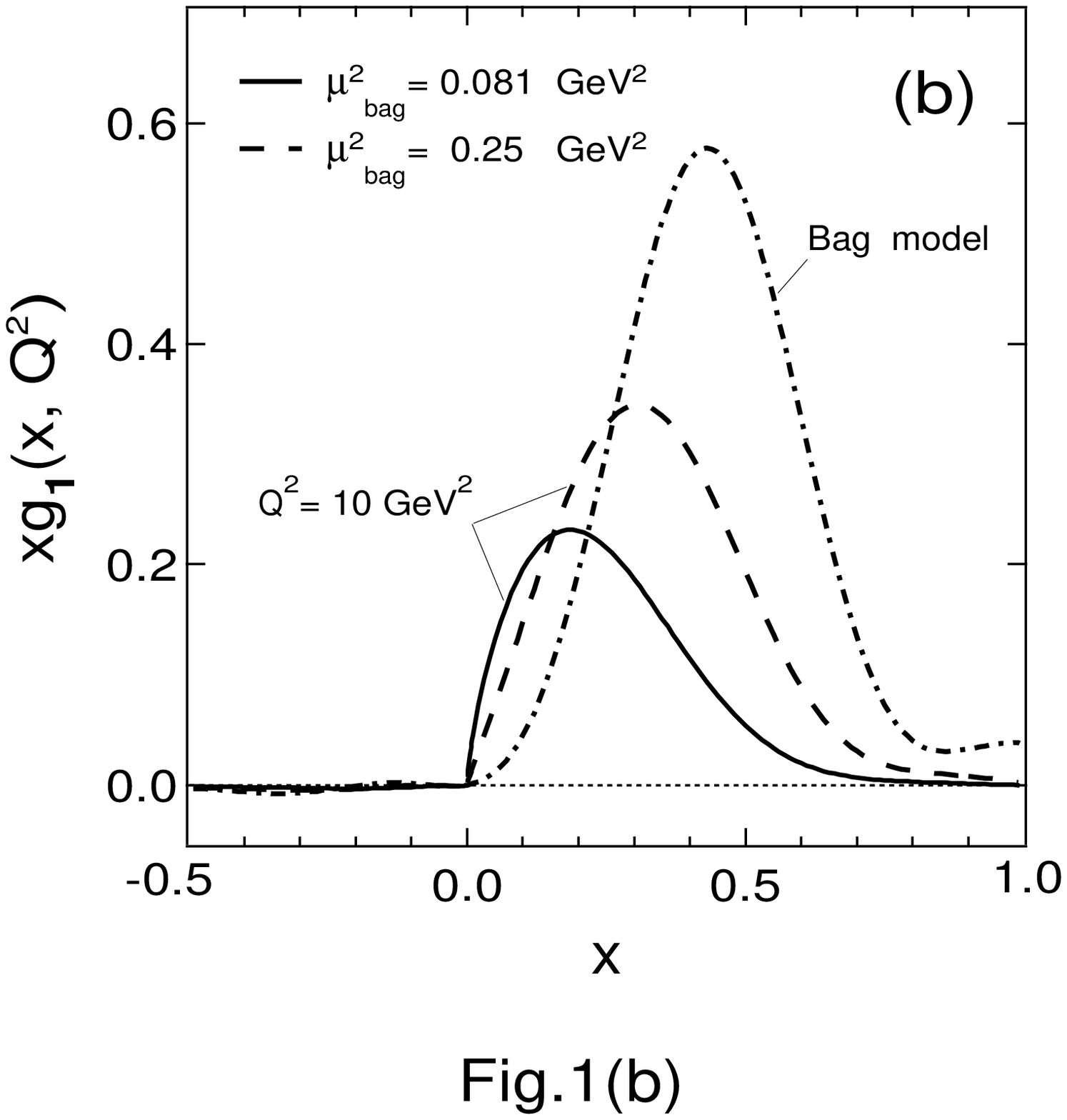}
\end{figure}

\begin{figure}[p]
\epsfbox{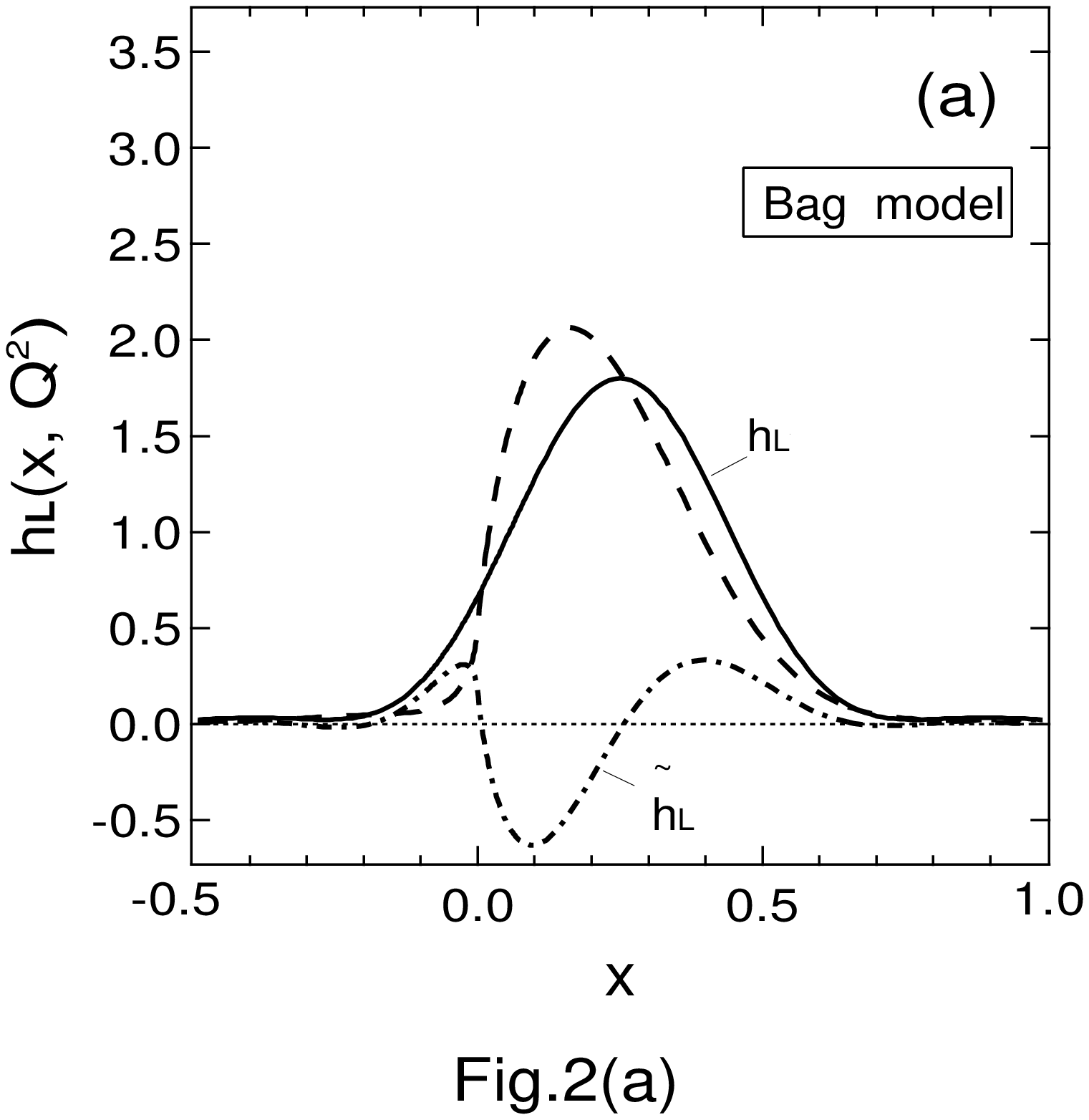}
\end{figure}

\begin{figure}[p]
\epsfbox{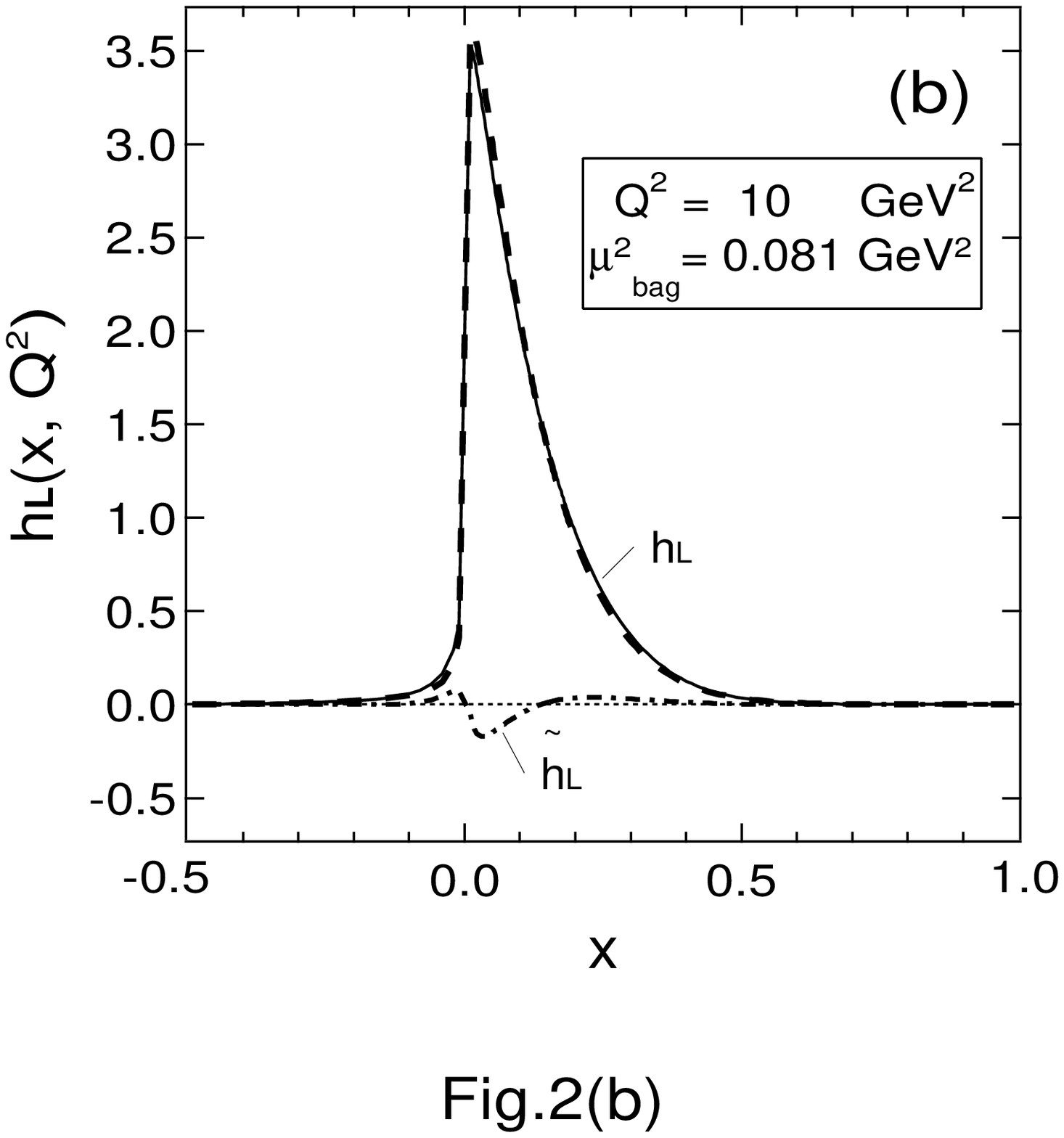}
\end{figure}

\begin{figure}[p]
\epsfbox{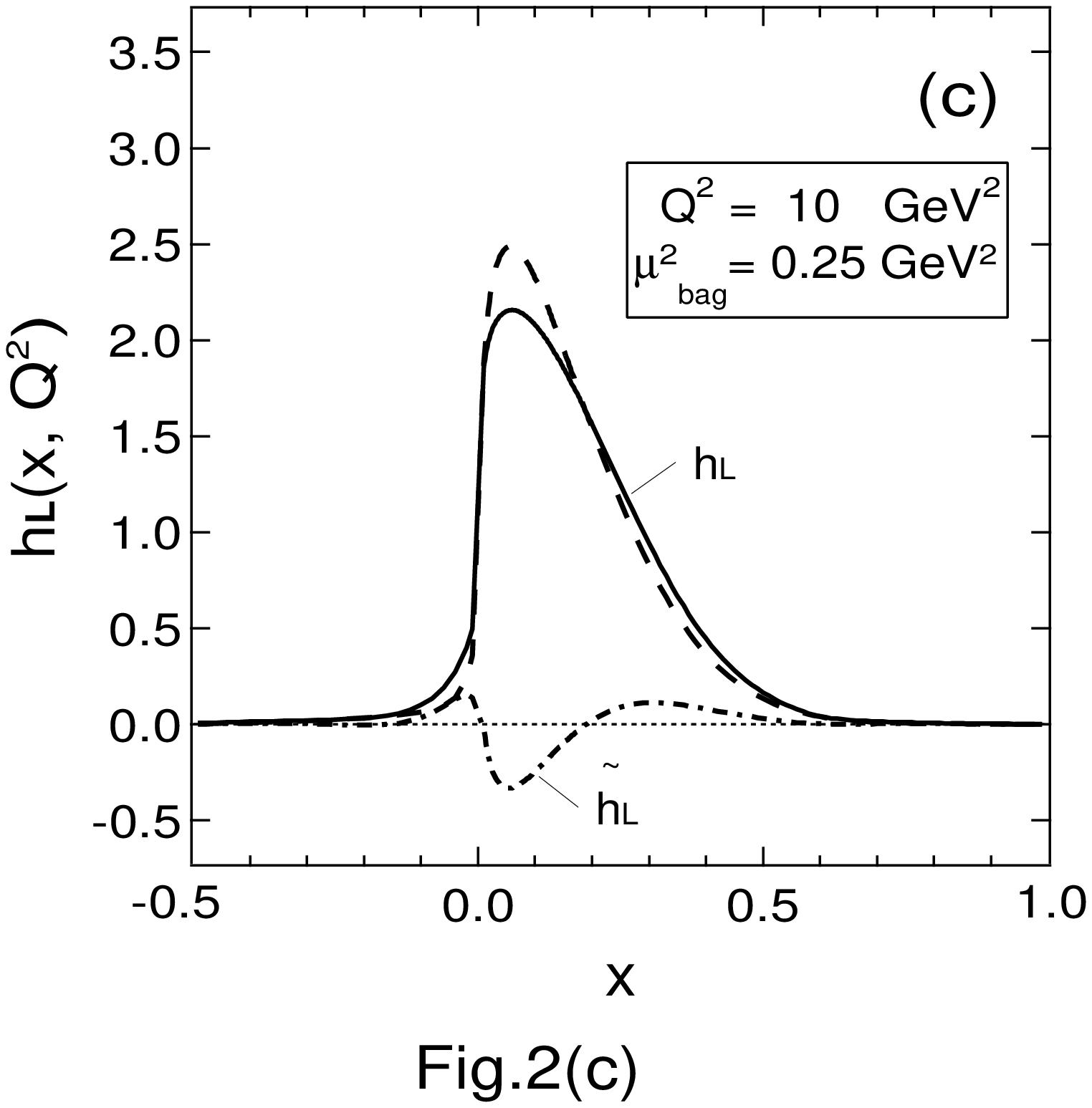}
\end{figure}

\begin{figure}[p]
\epsfbox{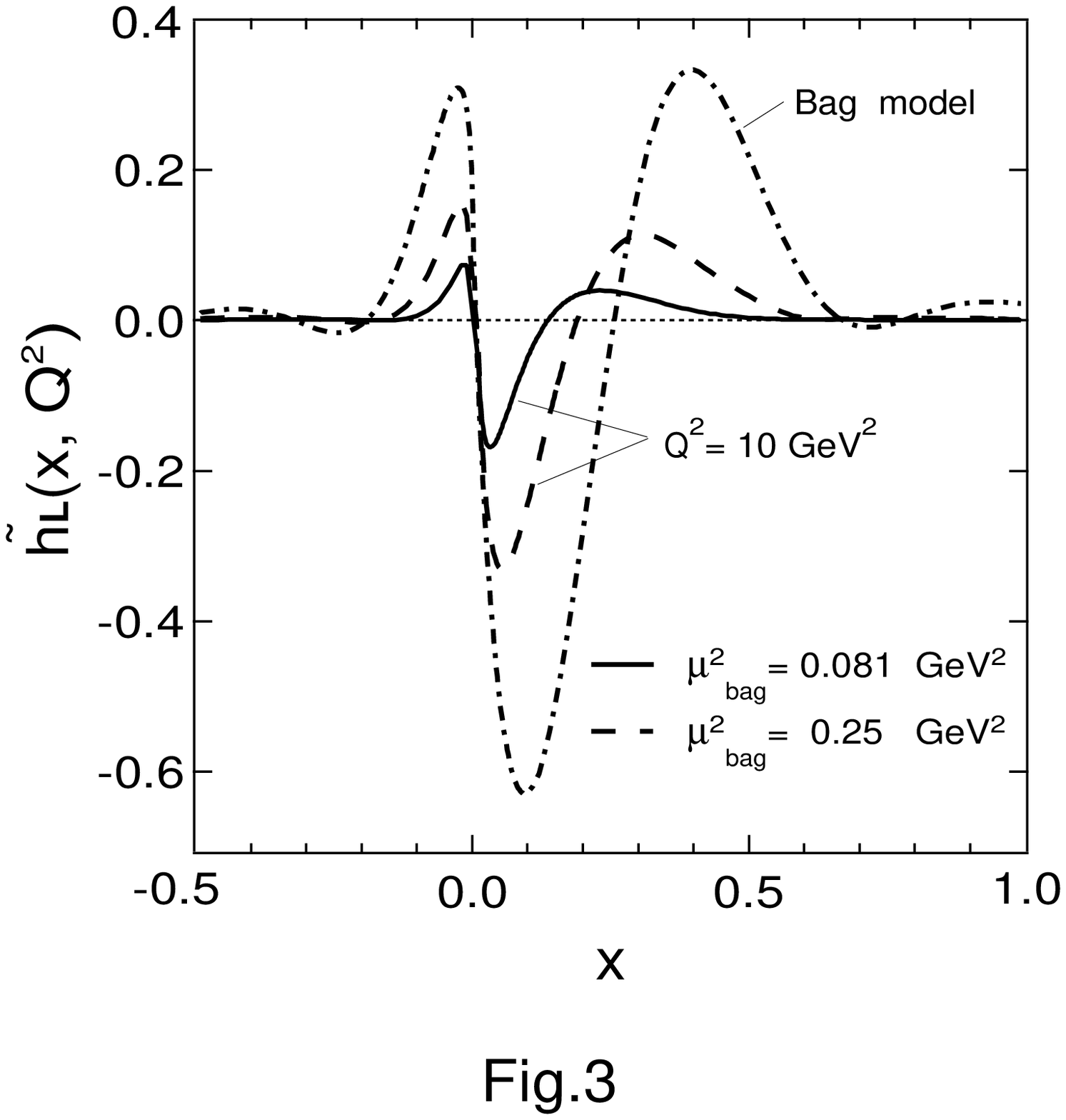}
\end{figure}

\begin{figure}[p]
\epsfbox{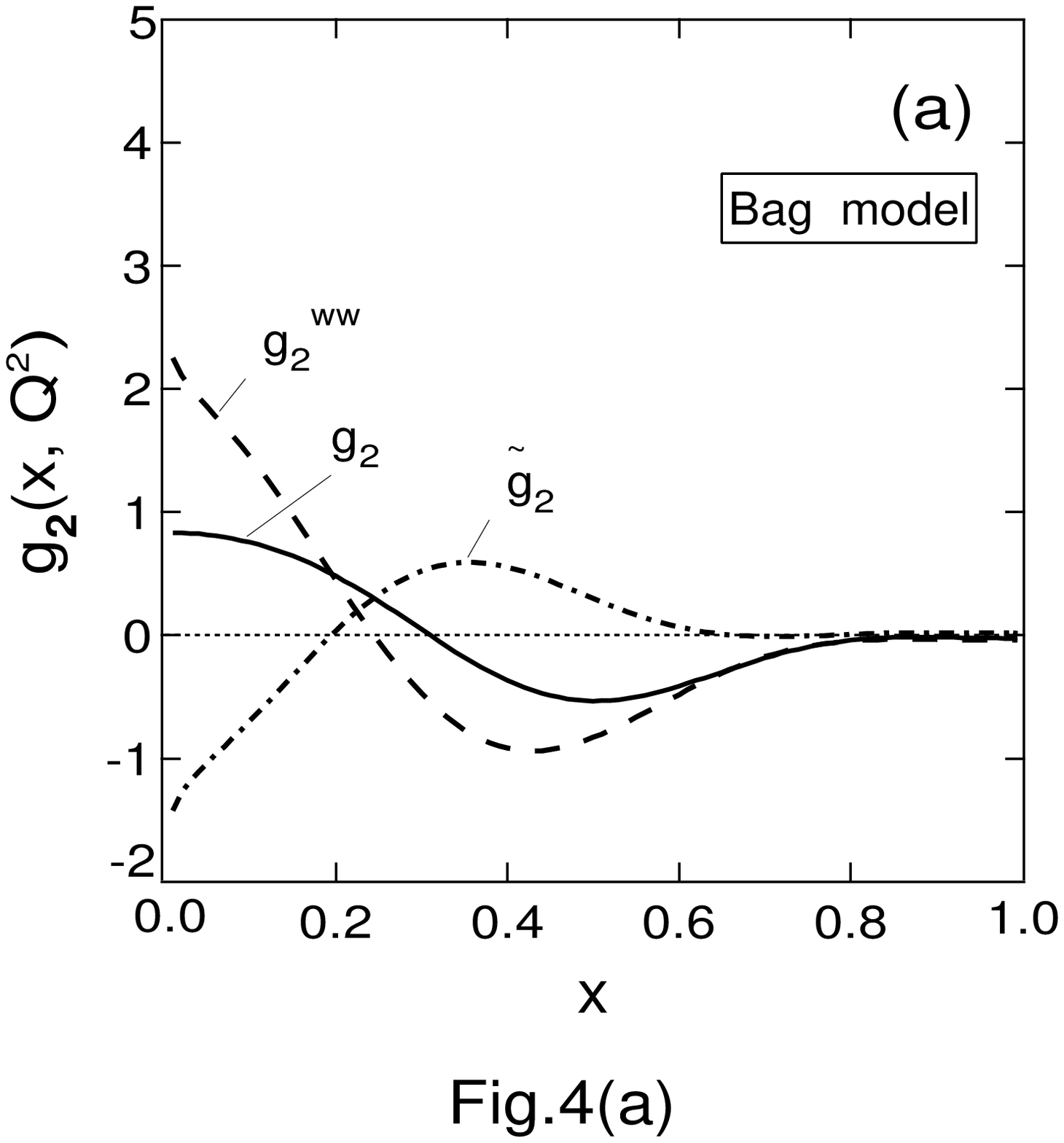}
\end{figure}

\begin{figure}[p]
\epsfbox{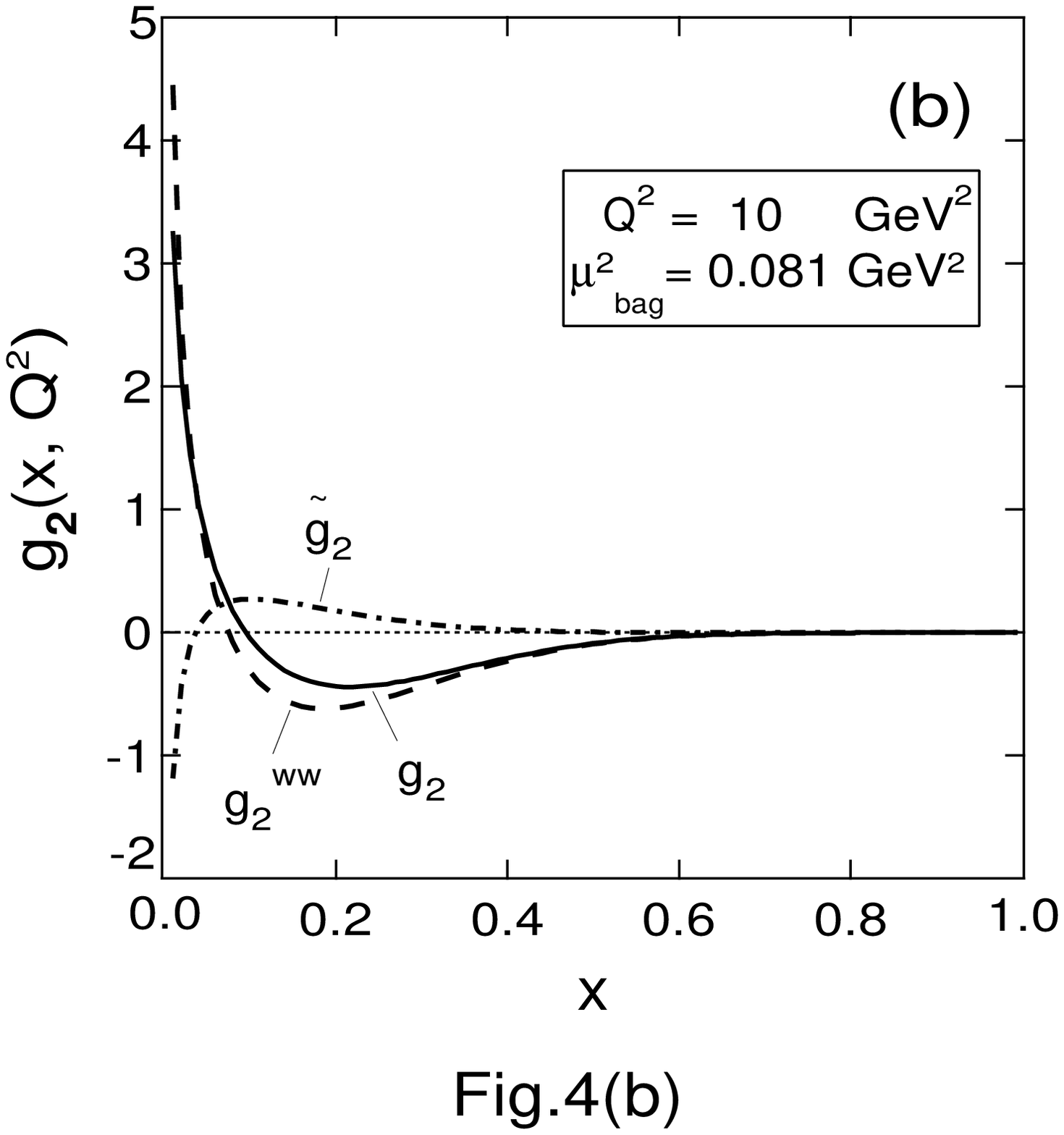}
\end{figure}

\begin{figure}[p]
\epsfbox{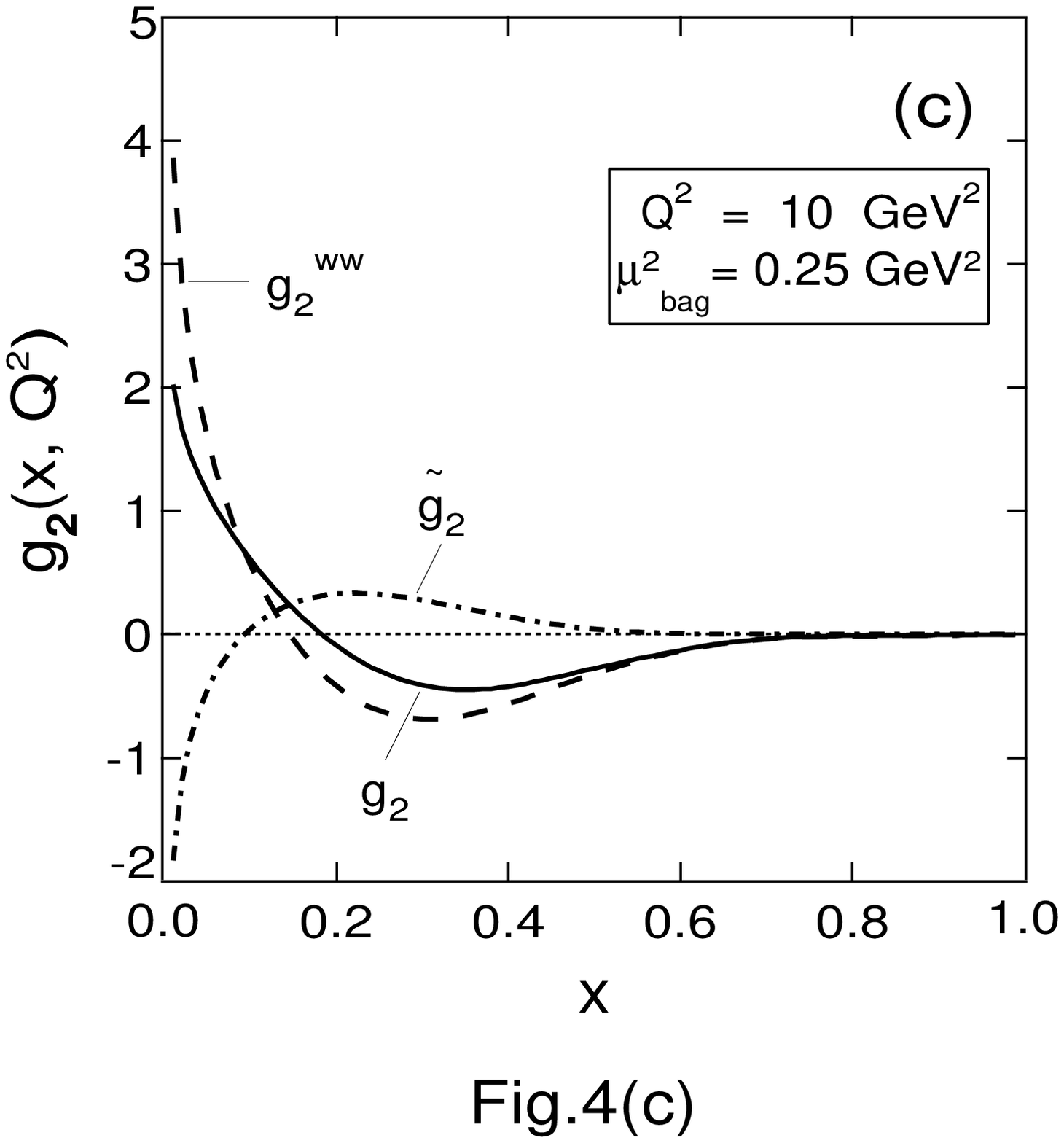}
\end{figure}

\end{document}